\documentclass{article}
\pdfoutput=1
\usepackage{arxiv}
\usepackage[utf8]{inputenc} 
\usepackage[T1]{fontenc}    
\usepackage{hyperref}       
\hypersetup{ colorlinks=true, linkcolor=blue, filecolor=magenta, urlcolor=cyan, }

\usepackage{url}            

\usepackage{amsmath,amssymb,amsfonts}
\usepackage{algorithmic}
\usepackage{graphicx}
\usepackage{textcomp}
\usepackage{filecontents,lipsum}
\usepackage[noadjust]{cite} 
\usepackage{times}
\usepackage{graphicx}
\usepackage{amsmath}
\usepackage{amssymb}
\usepackage[ruled,vlined]{algorithm2e}
\usepackage{multirow}
\usepackage{booktabs}
\usepackage{adjustbox}
\usepackage{caption}
\usepackage{subcaption}
\usepackage{tabularx}
\usepackage{upgreek}
\usepackage{dsfont}

\DeclareCaptionFont{9pt}{\fontsize{9pt}{10pt}\selectfont}
\begin{document}
\title{Synthesis of Brain Tumor MR Images for Learning Data Augmentation}

\author{Sunho Kim, Byungjai Kim, and HyunWook Park,
\thanks{This work was supported by Institute for Information $\&$ communications Technology Promotion (IITP) grant funded by the Korea government (MSIT) (N01190056, Development of Explainable Human-level Deep Machine Learning Inference Framework).}
\thanks{Sunho Kim, Byungjai Kim, and HyunWook Park are with the School of Electrical Engineering, Korea Advanced Institute of Science and Technology (KAIST), Daejeon, Korea (e-mails: ksunho0660@kaist.ac.kr, byungjai05@gmail.com, and hwpark@kaist.ac.kr).}}

\maketitle

\begin{abstract}
Medical image analysis using deep neural networks has been actively studied. Deep neural networks are trained by learning data. For accurate training of deep neural networks, the learning data should be sufficient, of good quality, and should have a generalized property. However, in medical images, it is difficult to acquire sufficient patient data because of the difficulty of patient recruitment, the burden of annotation of lesions by experts, and the invasion of patients’ privacy. In comparison, the medical images of healthy volunteers can be easily acquired. Using healthy brain images, the proposed method synthesizes multi-contrast magnetic resonance images of brain tumors. Because tumors have complex features, the proposed method simplifies them into concentric circles that are easily controllable. Then it converts the concentric circles into various realistic shapes of tumors through deep neural networks. Because numerous healthy brain images are easily available, our method can synthesize a huge number of the brain tumor images with various concentric circles. We performed qualitative and quantitative analysis to assess the usefulness of augmented data from the proposed method. Intuitive and interesting experimental results are available online at \url{https://github.com/KSH0660/BrainTumor}
\end{abstract}

\section{Introduction} 

Medical image analysis technologies have evolved considerably thanks to emerging deep neural networks. In general, well-annotated labels are required for supervised learning of deep neural networks. Several institutions provide public datasets of patient brain MRI images, such as BraTS \cite{menze2014multimodal}, ADNI, and ISLES. These datasets have good quality and easy accessibility, but the quantity of data might be insufficient for the learning of deep neural networks. It is very difficult to obtain a large amount of patient data, which is costly and time-consuming. In comparison, obtaining the medical data of healthy volunteers is relatively easy. In addition, there are many medical datasets obtained from healthy people, such as HCP \cite{van2013wu} and OASIS.

The most common solutions to overcome the lack of patient data are data augmentation methods, such as flip, rotate, and color jittering. These methods are simple to implement and widely used, and they often yield decent performance gains. Variations of the position, angle, brightness, and contrast of an image can be efficient for data augmentation, but this approach has limitations in terms of sufficient diversity \cite{eaton2018improving}.

To ensure diversity, there are several algorithms \cite{radford2015unsupervised, arjovsky2017wasserstein, karras2017progressive, mok2018learning,shin2018medical} for the synthesis of brain tumor images using a generative model, especially generative adversarial networks (GANs) \cite{goodfellow2014generative}. Several GAN-based algorithms \cite{radford2015unsupervised, arjovsky2017wasserstein, karras2017progressive} use random noise vectors (i.e. unconditional GAN) as input data, and synthesize brain tumor images. However, unintended images could be generated by the unconditional GAN algorithms. Other algorithms \cite{mok2018learning,shin2018medical} use tumor masks rather than noise vectors as the input data for tumor image synthesis. They successfully synthesize brain tumor images as intended. However, because simple modification of tumor masks is done as the input condition, the synthesized images are not significantly different from the existing training dataset.

The proposed method synthesizes brain tumor images from normal brain images and concentric circles that are simplified tumor masks. The tumor masks are defined by complex features, such as grade, appearance, size, and location. Thus, these features of the tumor masks are condensed and simplified to concentric circles. In the proposed method, the user-defined concentric circles are converted to various tumor masks through deep neural networks. The normal brain images are masked by the tumor mask, and the masked region is inpainted with the tumor images synthesized by the deep neural networks. Because the non-tumor parts are not synthesized but filled with real normal brain images, the synthesized brain tumor images have a realistic appearance. In addition, in terms of data augmentation, a large number of tumor images can be generated because the number of normal images is much larger than that of patient images as shown in Table \ref{table:tabel_comparison_patient&normal}.

\begin{table}[ht!]
\centering
\caption{Patient's dataset and normal dataset}
\label{table:tabel_comparison_patient&normal}
\begin{adjustbox}{max width=\linewidth} 
\begin{tabular}{@{}crl@{}}
\toprule[1pt]\midrule[0.3pt]
                                                                              & Medical image datasets & Data amounts     \\ \midrule
\multirow{3}{*}{Patient}                                                      & BraTS                  & 285 patients     \\
                                                                              & KiTS                   & 210 patients     \\
                                                                              & LiTS                   & 130 patients     \\ \midrule[0.3pt]
\multirow{2}{*}{Normal} & HCP                    & 1206 individuals \\
                                                                              & OASIS                  & 1098 individuals \\  \midrule[0.3pt]\bottomrule[1pt]
                                                                              
\end{tabular}
\end{adjustbox}
\end{table}

The remaining sections are organized as follows.
Section \ref{sec:previous works} introduces previous works about the synthesis of medical images using GAN. Section \ref{sec:method} presents our proposed method, its training scheme, and various loss functions and implementation details.
In Section \ref{sec:experiments}, extensive experiments to evaluate
the proposed method are presented. Some issues are discussed in Section \ref{sec:discussion}. Finally,
our conclusions are presented in Section \ref{sec:conclusion}.

\section{Related Work}
\label{sec:previous works}

\subsection{Medical Image Synthesis with Generative Networks}
Tumor segmentation methods \cite{ronneberger2015u, pereira2016brain, havaei2017brain} have used conventional augmentation methods such as shift and rotation \cite{ronneberger2015u}, patch rotation \cite{pereira2016brain}, and flip \cite{havaei2017brain}, to improve the learning performance of the deep neural network. However, the diversity of data augmented by conventional methods is limited. To increase the diversity of data, previous works have used the GAN-based data augmentation methods \cite{mok2018learning, shin2018medical, bermudez2018learning, han2018gan, bowles2018gan}. In synthesizing the brain tumor magnetic resonance (MR) images, not only images but also tumor masks must be generated. This is because both images and tumor masks are used for supervised learning in the brain tumor segmentation. Conditional synthesis methods \cite{mok2018learning, shin2018medical} synthesize brain tumor images as well as tumor masks, whereas unconditional synthesis methods \cite{bermudez2018learning, han2018gan, bowles2018gan} synthesize only the brain tumor images.

Several GAN methods (e.g. DCGAN \cite{radford2015unsupervised}, WGAN \cite{arjovsky2017wasserstein}, PGGAN \cite{karras2017progressive}) have been studied for unconditional synthesis. They usually generate images from random noise. The learning phase may be simple, but there are limitations to generating the intended output.
DCGAN is a basic and simple GAN method. Bermudez \textit{et al.} \cite{bermudez2018learning} proposed a method that  synthesizes brain tumor images using DCGAN. WGAN uses another optimization method, and generally produces better output performance; it was used to synthesize brain tumor images by Han \textit{et al.} \cite{han2018gan}. PGGAN is a progressive learning scheme for the best performance. Christopher Bowles \textit{et al.} \cite{bowles2018gan} proposed the method that synthesizes of brain tumor images by a PGGAN-based method. However, all of these methods have the disadvantage that they cannot be used for supervised learning because they are unconditional synthesis methods (i.e. only tumor images are synthesized without tumor masks).

The conditional synthesis methods do not generate the images from random noise, but from a brain tumor label map, where the brain tumor label map consists of a brain image and a tumor mask. They can be used as augmentation methods for supervised learning about tumor segmentation. Shin \textit{et al.} \cite{shin2018medical} proposed a method that uses a pix2pix \cite{isola2017image} model. This method synthesizes brain tumor MR images from brain atlas maps. Mok and Chung \cite{mok2018learning} proposed a coarse-to-fine GAN to synthesize brain tumor MR images using deformed label maps of 2D axial slices as conditional inputs. However, because the label map transformed from an existing one is used as the conditional input, the number of images that can be synthesized is limited and the results may be unrealistic. 
\begin{figure*}[ht!]
\centering
  \includegraphics[width=\linewidth]{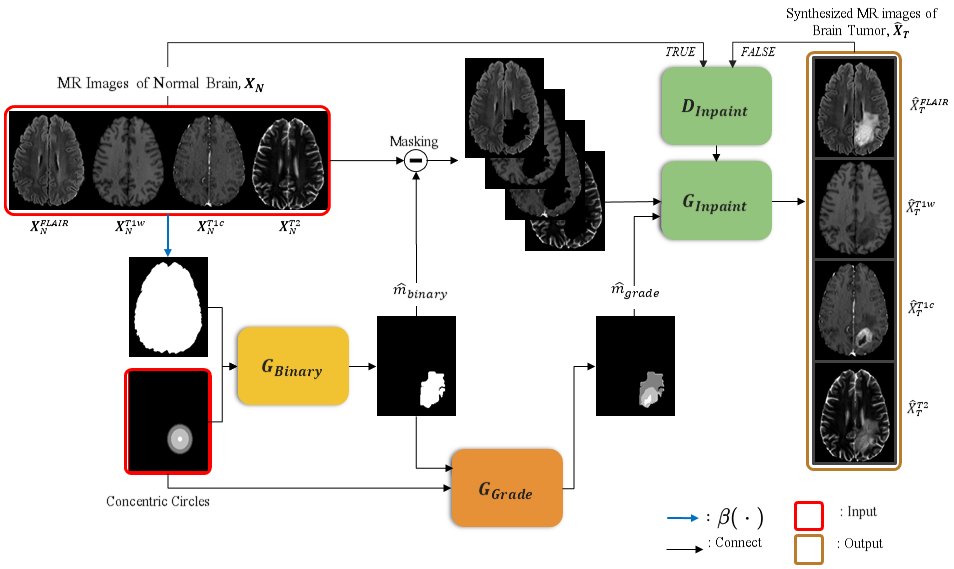}
  \caption{Overall structure of the proposed method for synthesis of brain tumor images. Normal brain images and concentric circles are inputs of the overall networks. Blue arrows represent the binary operation, $\beta(\cdot)$. Black arrows represent data flows.}
  \label{fig:test}
\end{figure*}

\begin{figure}[ht!]
\centering
  \includegraphics[width=0.5\linewidth]{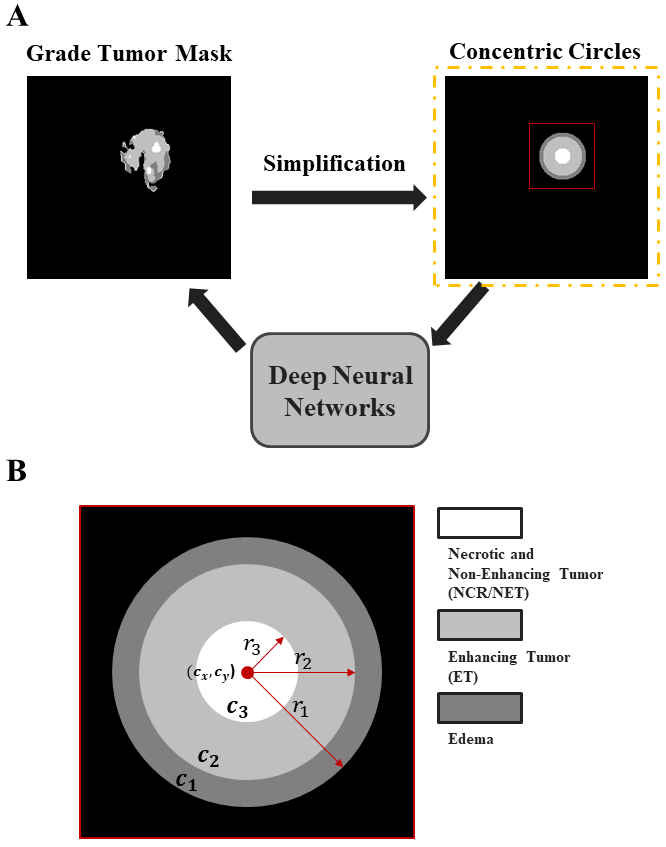}
  \caption{(A) Simplification of grade tumor mask into concentric circles. As reverse process of the simplification, concentric circles are converted to grade tumor masks through deep neural networks. (B) There are five variables defining the concentric circles; coordinates of the center, $(c_x, c_y)$, and radii of the three concentric circles, $r_1, r_2$, and $r_3$.}
  \label{fig:intro}
\end{figure}

\subsection{User-guided Image Inpainting through Deep Learning}
\label{sec:User-guided Image Inpainting with Deep Learning}
There are several inpainting methods to fill holes in images using deep learning \cite{yu2018generative,liu2018image, iizuka2017globally, yeh2017semantic, zhao2019guided, yu2019free}. The holes are filled with the contextual attention of the background \cite{yu2018generative}. It uses coarse-to-fine networks to fill the hole more harmoniously. Partial convolution is used to fill the holes in an image gradually from the outside \cite{liu2018image}. Both global and local discriminators are used to harmonize the filled hole and the entire image \cite{iizuka2017globally}. The image inpainting was performed by finding variables with which the holes in an image can be inpainted well \cite{yeh2017semantic}. The methods mentioned above inpaint the hole in an image by using background. 

There are several ways to inpaint images by user-guidance \cite{zhao2019guided, yu2019free}. A hole can be inpainted by using a synthesized image \cite{zhao2019guided}. A free-form mask is filled by a user-guided sketch \cite{yu2019free}. A given image was transformed into various images using guidance masks \cite{park2019semantic}. The network of \cite{park2019semantic} used the spatially-adaptive normalization of the feature maps for mask-dependency. 

\section{The Proposed Method}
\label{sec:method}

As shown in Fig.\ref{fig:test}, the proposed method synthesizes multi-contrast brain tumor MR images using two kinds of inputs, namely, multi-contrast normal brain MR images and concentric circles. It consists of four networks; two networks of $G_{binary}$ and $G_{grade}$ for generation of the tumor mask, and two network of $G_{inpaint}$ and $D_{inpaint}$ for synthesis of the brain tumor images. At first, $G_{binary}$ generates a binary tumor mask, $\hat{m}_{binary}$, which represents the tumor's geometric features, that is, the size, position, and appearance. Second, $G_{grade}$ generates the grade of the tumor binary mask, whose output is the grade tumor mask, $\hat{m}_{grade}$. Finally, $G_{inpaint}$ synthesizes the multi-contrast brain tumor MR images, and $D_{inpaint}$ is the discriminator network to evaluate $G_{inpaint}$.

Let $\textbf{X}_T= \left \{X_T^{FLAIR}, X_T^{T1w},X_T^{T1c},X_T^{T2w} \right \}$ denote the multi-contrast MR images of the brain tumor, and let $ m_{grade}$ be the grade tumor mask. The superscripts of FLAIR, T1w, T1c, and T2w denote images obtained by fluid attenuated inversion recovery (FLAIR), and T1-weighted, contrast-enhanced T1-weighted, and T2-weighted imaging sequences, respectively. The multi-contrast normal brain images are denoted by $\textbf{X}_N=\left \{X_N^{FLAIR}, X_N^{T1w},X_N^{T1c},X_N^{T2w} \right \}$. We use an operator, $\beta(\cdot),$ for binary masking of objects. For example, $\beta(m_{grade})$ means the binarized grade tumor mask, i.e. binary tumor mask, $m_{binary}$. 

\subsection{Generation of the Grade Tumor Masks}
\label{sec:Generation of the Grade Tumor Masks}

There are three grades of brain tumors, as shown in Fig.\ref{fig:intro}. The geometric features and grade information of tumors can be simplified to concentric circles. The concentric circles, $c=\{c_1, c_2, c_3\}$, are generated from the tumor masks according to two policies. The three concentric circles should have the same center position as the given tumor mask and the area per grade of tumor should be equal to the area of $c_1-c_2, c_2-c_3,$ and $c_3,$ which correspond to the area of edema, enhancing tumor, and necrotic and non-enhancing tumor, respectively, as shown in Fig.\ref{fig:intro}. As shown in Fig.\ref{fig:intro}A, the grade tumor masks are simplified into concentric circles by the aforementioned policies. There are five variables that define the concentric circles, i.e., coordinates of the center of the concentric circles, ($c_x, c_y$), and the radii of three concentric circles, $r_1,\ r_2$, and $r_3,$ which represent the grade information as shown in Fig.\ref{fig:intro}B.  Therefore, a new tumor mask can be simply generated from these five variables.

Then, two networks, $G_{bianry}$ and $G_{grade}$, convert the concentric circles to grade tumor masks. For conversion from the concentric circles to the grade tumor mask, initially $G_{binary}$ generates the binary tumor mask, $\hat{m}_{binary}$, as follows:
\begin{equation}
\begin{aligned}
\hat{m}_{binary} = G_{binary}(c_1,\ \beta(X_N^{T1w}))
\end{aligned}
\end{equation}
where $c_1$ is the outermost concentric circle, which represents the whole tumor region, and $\beta(\cdot)$ means the binarization operator. Thus, $\beta(X_N^{T1w})$ represents the binarized T1-weighted image of the normal brain, i.e. the shape of the normal brain. Any contrast MR image of the normal brain can be used for this process. At this step, the geometric shape of the tumor is produced from the concentric circles and the binarized normal brain image. The appearance is determined by the appropriate combination of the brain's shape and the concentric circles. A detailed description of the appropriate combination will be explained in section \ref{sec:experiments}.

\begin{figure*}[ht!]
\centering
  \includegraphics[width=0.7\linewidth]{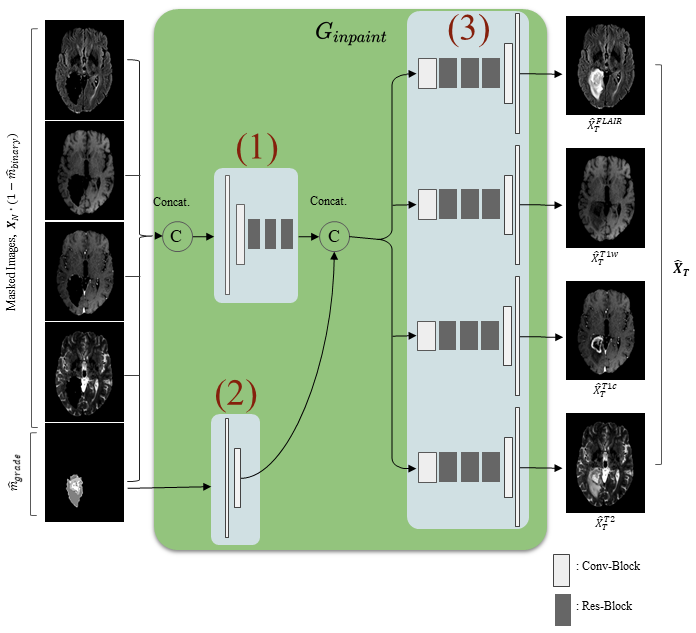}
  \caption{The proposed network architecture of $G_{inpaint}$. Five channels of concatenated images are forwarded to (1) and the tumor mask is forwarded to (2). The feature maps of outputs of (1) and (2) are concatenated and then forwarded to each branch network of (3). White and grey blocks represent CNN-based blocks and residual-based blocks, respectively. $\textcopyright$ means concatenation.}
  \label{fig:inpaint-net}
\end{figure*}

After generation of the binary tumor mask, $\hat{m}_{binary}$, $G_{grade}$ generates the grade tumor mask, $\hat{m}_{grade}$, as follows: 
\begin{equation}
\begin{aligned}
\hat{m}_{grade}=G_{grade}(c_1, c_2, c_3,\ \hat{m}_{binary})
\end{aligned}
\end{equation}
Here, $G_{grade}$  aims to determine the grade information of $\hat{m}_{binary}$, by referring the area of $c_1-c_2, c_2-c_3,$ and $c_3$. Through this step, concentric circles are converted to the grade tumor mask. The ability to generate appropriate grade tumor masks from easily controllable concentric circles has great advantages in terms of the diversity of the augmented data.

\subsection{Synthesis of Brain Tumor Images }
\label{sec:synthesis of brain tumor images}

Brain tumor images are synthesized using normal brain images, $\textbf{X}_N$, and a grade tumor mask, $\hat{m}_{grade}$, through $G_{inpaint}$. As a preliminary procedure of $G_{inpaint}$, the tumor parts of the normal brain images which are the same region as $\hat{m}_{grade}$ are masked. The masked region are inpainted with the synthesized tumor through $G_{inpaint}$, as follows:
\begin{equation}
\begin{aligned}
\hat{\textbf{X}}_{T}=G_{inpaint}\left ( \textbf{X}_N\cdot(1-\hat{m}_{binary}), \ \hat{m}_{grade} \right ),
\end{aligned}
\end{equation}where the multiplication factor of $(1-\hat{m}_{binary})$ represents the masking process. Here, $G_{inpaint}$ harmonically synthesizes the tumor-part images with the non-tumor parts. Also, $\hat{m}_{grade}$ is used in $G_{inpaint}$ for user-guided image inpainting as mentioned in section \ref{sec:User-guided Image Inpainting with Deep Learning}. As shown in Fig.3, $G_{inpaint}$ utilizes masked multi-contrast images and $\hat{m}_{grade}$ as input images and generates multi-contrast tumor images.

\begin{algorithm}[ht!]
\SetAlgoLined
j = 0\;
\While{j $\leq$ N}{
$\textbf{X}_N=\left \{X_N^{FLAIR}, X_N^{T1w},X_N^{T1c},X_N^{T2w} \right \}$\;
Generate two random, $c_x, c_y\ \in \ [80,\ 160] $\;
Generate three random, $r_1, r_2, r_3\ \in \ \textit{sorted}[0,\ 40]$\;
The concentric circles, $c_1, c_2, c_3$ are determined by five variables, $c_x, c_y, r_1, r_2, r_3$\;
    $\hat{m}_{binary} = G_{binary}(c_1,\ \beta(X_N^{T1w}))$\;
    $\hat{m}_{grade}=G_{grade}(c_1,\ c_2,\  c_3,\ \hat{m}_{binary})$\;
    \eIf{$\hat{m}_{binary} \cdot (1-\beta(X_N^{T1w})) == 0$}{
    $\hat{\textbf{X}}_{T}=G_{inpaint}\left( \textbf{X}_N\cdot(1-\hat{m}_{binary}), \ \hat{m}_{grade} \right )$\;
   j $\gets$ j+1\;
   }{j $\gets$ j\;}
 }
 \caption{How to synthesize brain tumor MR images (Number of images is \textit{N})}
\end{algorithm}

The steps for synthesizing a total of N brain tumor images are described in Algorithm 1. The concentric circles are generated by five randomly selected variables, where the generated tumor masks should be located within the brain. This simple and reliable algorithm can quickly synthesize brain tumor images.

\begin{figure}[ht!]
\centering
  \includegraphics[width=0.8\linewidth]{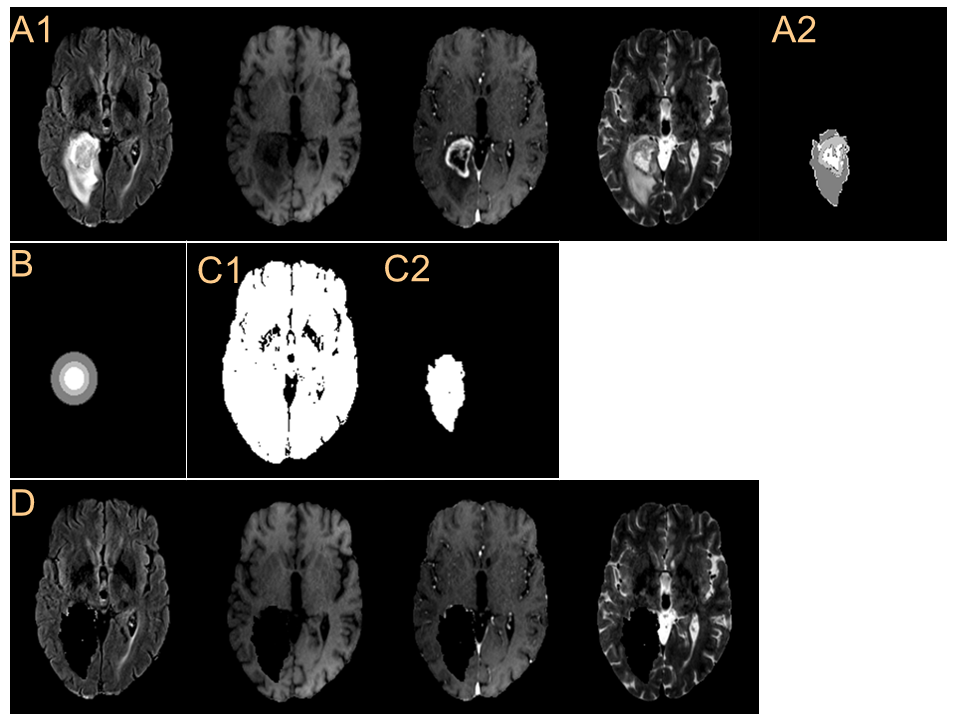}
  \caption{Data preprocessing for training of the proposed method. (A1) Multi-contrast brain tumor MR images, $\textbf{X}_{T}$, and (A2) grade tumor mask, $m_{grade}$ from BraTS 2018. (B) Concentric circles, $c_1, c_2,$ and $c_3$, simplified from $m_{grade}$. (C1) Shape of the brain and (C2) binary tumor mask, $m_{binary}$ are produced by binarization operator, $\beta(\cdot)$. (D) Multi-contrast brain images are masked in the tumor-parts of $\textbf{X}_{T}$.}
  \label{fig:train}
\end{figure}

\subsection{Network Training and Loss Functions}
\label{sec:learning and loss functions}

BraTS 2018 \cite{menze2014multimodal} is used as the dataset for training of the proposed network. For inputs and ground truths of each network, the given images from the dataset are preprocessed as shown in Fig.\ref{fig:train}. There are brain tumor MR images, $\textbf{X}_T,$ and grade tumor mask, $m_{grade},$ as shown in Fig.\ref{fig:train}(A1-A2). The concentric circles, $c_1, c_2,$ and $c_3,$ by simplifying the tumor mask are shown in Fig.\ref{fig:train}(B). The shapes of the brain and $m_{binary}$ are shown in Fig.\ref{fig:train}(C1-C2). Masked brain tumor images are produced by multiplication of a factor of $(1-m_{binary})$ as shown in Fig.\ref{fig:train}(D). The proposed networks are trained by using preprocessed images and given images from the dataset, as input and ground truths for each network, such as $m_{binary}$ and $m_{grade}$, not $\hat{m}_{binary}$ and $\hat{m}_{grade}$. Details of the loss function used in each training will be explained. 

First, $G_{binary}$ is trained to generate $m_{binary}$ using $c_1$ and $\beta(X_T^{T1w})$ through minimization of pixel loss, $\mathcal{L}_{binary}$, as follows:
\begin{equation}
\begin{aligned}
 \mathcal{L}_{binary} =\left \| m_{binary}-G_{binary}\left (c_1,\beta(X_T^{T1w})\right )\right \|_1\ 
\end{aligned}
\end{equation}
where $c_1$ is the outer-most concentric circle, $\beta(X_T^{T1w})$ represents the shape of the brain, and $\left \| \cdot \right \|_1$ denotes the absolute sum.

Similarly, $G_{grade}$ is trained to estimate $m_{grade}$ from the concentric circles and $\hat{m}_{binary}$ through minimization of the pixel loss, $\mathcal{L}_{grade}$, as follows:
\begin{equation}
\begin{aligned}
\mathcal{L}_{grade}={\displaystyle \left \| m_{grade}-G_{grade}\left (c_1, c_2, c_3, m_{binary} \right ) \right \|_1}
\end{aligned}
\end{equation}

For the synthesis of brain tumor images, various loss functions are used. The pixel loss function, $\mathcal{L}_{inpaint:pix}$; a content loss function, $\mathcal{L}_{inpaint:cont}$; and an adversarial loss function, $\mathcal{L}_{inpaint:adv}$, are used for the training of $G_{inpaint}$ and $D_{inpaint}$, where $D_{inpaint}$ is the discriminator network. The pixel loss function, $\mathcal{L}_{inpaint:pix}$, is defined as follows:
\begin{equation}
\begin{aligned}
&\mathcal{L}_{inpaint:pix}\\
& = {\displaystyle  \left \| \textbf{X}_T-G_{inpaint}\left ( \textbf{X}_T \cdot (1-m_{binary}), m_{grade}\right ) \right \|_1}
\end{aligned}
\end{equation}

If only $\mathcal{L}_{inpaint:pix}$ is used for training, it is not sufficient to make the output a realistic tumor in shape. The content loss function can make the output more perceptually similar to the ground truth, which is defined as follows:
\begin{equation}
\begin{aligned}
&\mathcal{L}_{inpaint:cont}\\
&={\displaystyle \left \| \Psi (\textbf{X}_T)- \Psi (G_{inpaint}\left ( \textbf{X}_T \cdot (1-m_{binary}), m_{grade}\right ) \right \|_1},
\end{aligned}
\end{equation}where $\Psi$ represents the second layer's output of VGG-19 \cite{simonyan2014very}. This loss function is commonly used to match image styles \cite{gatys2016image,johnson2016perceptual}, which is also known as a perceptual loss. 

The adversarial learning scheme makes the output more realistic and sharp in texture. $\mathcal{L}_{inpaint:adv}$ for  $G_{inpaint}$ and $D_{inpaint}$ is described as
\begin{equation}
\begin{aligned}
&\mathcal{L}_{inpaint:adv}={\mathbb{E}_{\textbf{x}\sim \mathbb{P}_R}}[D_{inpaint}(\textbf{x}, m_{grade})]+{\mathbb{E}_{\hat{\textbf{x}}\sim \mathbb{P}_G}}[1-\\
&D_{inpaint}(G_{inpaint}\left (\textbf{X}_T \cdot (1-m_{binary}), m_{grade}\right ), m_{grade}))],\
\end{aligned}
\end{equation}where $\mathbb{E}_{\textbf{x}\sim \mathbb{P}_R}$ and $\mathbb{E}_{\hat{\textbf{x}}\sim \mathbb{P}_G}$ are the expectation values over the real brain tumor images and the synthesized brain tumor images, respectively. Here, $G_{inpaint}$ is trained to minimize $\mathcal{L}_{inpaint:adv}$, whereas $D_{inpaint}$ is trained to maximize $\mathcal{L}_{inpaint:adv}$.

While training of $G_{binary}$ and $G_{grade}$ uses only simple pixel loss function, training of $G_{inpaint}$ and $D_{inpaint}$ uses various loss functions to prevent the blurry and inapparent texture of outputs. A detailed analysis according to loss functions is presented in section \ref{sec:experiments}.

\subsection{Training Configurations and Implementation Details}

There are a total of four networks in the proposed method: $G_{binary}$, $G_{grade}$ , $G_{inpaint}$, and $D_{inpaint}$; $G_{binary}$ and $G_{grade}$ use the U-NET \cite{ronneberger2015u}. The input layers of $G_{binary}$ and $G_{grade}$ are both two-channel images of (256, 256, 2), and the output layer is a one-channel image of (256, 256, 1). The structure of $G_{inpaint}$ is shown in Fig.\ref{fig:inpaint-net}. The input layers of $G_{inpaint}$ are the tumor grade mask, which has one channel, (256, 256, 1), and four-channel multi-contrast MR images of the masked normal brain, (256, 256, 4). Then, the output layer of  $G_{inpaint}$ is four-channel synthesized brain tumor images of (256, 256, 4). The detailed network structures are explained in Table \ref{table:network_architecture}.  

 \begin{table}[ht!]
\centering
\caption{Network architectures of $G_{inpaint}$ and $D_{inpaint}$. CIR(k,s) is a combination of CNN with a kernel size of k and stride of s, Instance Normalization \cite{ulyanov2016instance}, and ReLU. RES(k) is a residual block with a kernel size of k. UP(k) is a upsampling block with a factor of k. AVG(k) is a two-dimensional average pooling block with a kernel size of k. The output size is indicated by channel x height x width.}
\label{table:network_architecture}
\begin{adjustbox}{max width=\linewidth} 
\begin{tabular}{@{}cll@{}}
\toprule [1pt] \midrule [0.5pt]
Part & Type & Output Size     \\ \midrule
\multicolumn{3}{c}{$\textbf{G}_{inpaint}$}    \\ [2pt]
(1)  & CIR(3, 2)            & 32 x 128 x 128  \\
     & CIR(3, 2)            & 128 x 64 x 64   \\
     & RES(3)               & 128 x 64 x 64   \\
     & RES(3)               & 128 x 64 x 64   \\
     & RES(3)               & 128 x 64 x 64   \\ 
(2)  & CIR(3)               & 256 x 64 x 64   \\
     & RES(3)                & 256 x 64 x 64   \\
     & RES(3)               & 256 x 64 x 64   \\
     & RES(3)               & 256 x 64 x 64   \\
     & UP(2)                   & 256 x 128 x 128 \\
     & CIR(3, 1)            & 128 x 128 x 128 \\
     & UP(2)                   & 128 x 256 x 256 \\
     & CIR(3, 1)            & 64 x 256 x 256  \\
     & CIR(7, 1)            & 1 x 256 x 256   \\ 
(3)  & CIR(3, 2)            & 4 x 128 x 128   \\
     & CIR(3, 2)            & 16 x 64 x 64    \\ \midrule [0.6pt]
\multicolumn{3}{c}{$\textbf{D}_{inpaint}$}             \\ [2pt]
     & CIR(3, 2)            & 32 x 128 x 128  \\
     & CIR(3, 2)            & 64 x 64 x 64    \\
     & CIR(3, 1)            & 256 x 64 x 64   \\
     & RES(3)               & 256 x 64 x 64   \\
     & RES(3)               & 256 x 64 x 64   \\
     & RES(3)               & 256 x 64 x 64   \\
     & CIR(3, 1)            & 64 x 64 x 64    \\
     & CIR(3, 1)            & 1 x 64 x 64     \\
     & AVG(64)              & 1 x 1 x 1       \\ \midrule [0.5pt] \bottomrule[1pt] 
\end{tabular}
\end{adjustbox}
\end{table}

Our method was implemented in the PyTorch framework with PyCharm 2018.3.4 on the Ubuntu 16.04.3 platform. All experiments were conducted on a workstation equipped with TITAN Xp graphic card with 12 GB memory. In the training of all networks, ADAM \cite{kingma2014adam} was used for optimization with a learning rate of 0.001. We terminated the back-propagation after 200 epochs for preventing over-fitting. Our whole scheme including four networks required seven days for training. We saved all network parameters every 10 epochs, and the networks that had the lowest validation loss were chosen among the 20 saved models. 

\section{Experiments} 
\label{sec:experiments}
\subsection{Data Preprocessing} 

We used the BraTS 2018 as the brain tumor dataset. Twenty-eight subjects of a total of 285 subjects were used to measure the validation loss of the networks and the tumor segmentation performance. A total of 190 subjects with good quality among the remaining 257 subjects were selected for training of the networks. These 190 subjects were subjectively selected according to whether the tumors were well represented in the multi-contrast images. For each subject, we selected 66 slices from the whole brain volume, excluding the end parts of the brain. Therefore, 66 slices of a total of 190 subjects, that is 12540 slices, were used as training data. Each image was normalized using Gaussian normalization and clipping over the range of [-0.5, 5]. Then, for the grade tumor masks and concentric circles, pixels corresponding to edema (ED), enhancing tumor (ET), and necrotic/non-enhancing tumor (NCT/NET) of the tumor were set to values of 0.5, 0.75, and 1, respectively.

\subsection{Synthesis of Brain Tumor Images}

\begin{figure*}[ht!]
\centering
  \includegraphics[width=0.9\linewidth]{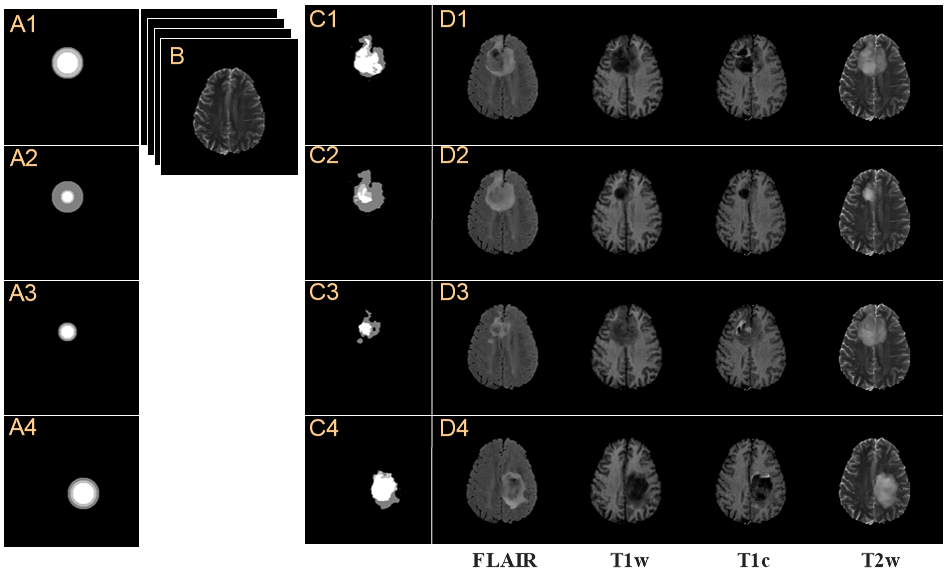}
  \caption{Synthesized brain tumor images, $\hat{\textbf{X}}_{T}$, from various concentric circles, $c$, with a given normal brain image, $\textbf{X}_N$. (A1-A4) Various concentric circles and (B) the given multi-contrast normal brain image, $\textbf{X}_N$. (C1-C4) Grade tumor masks, $\hat{m}_{grade}$, generated from various concentric circles. (D) Synthesized multi-contrast brain tumor images, $\hat{\textbf{X}}_{T}$.}
  \label{fig:exp_samebrain_variousmask}
\end{figure*}

\begin{figure*}[ht!]
\centering
  \includegraphics[width=0.9\linewidth]{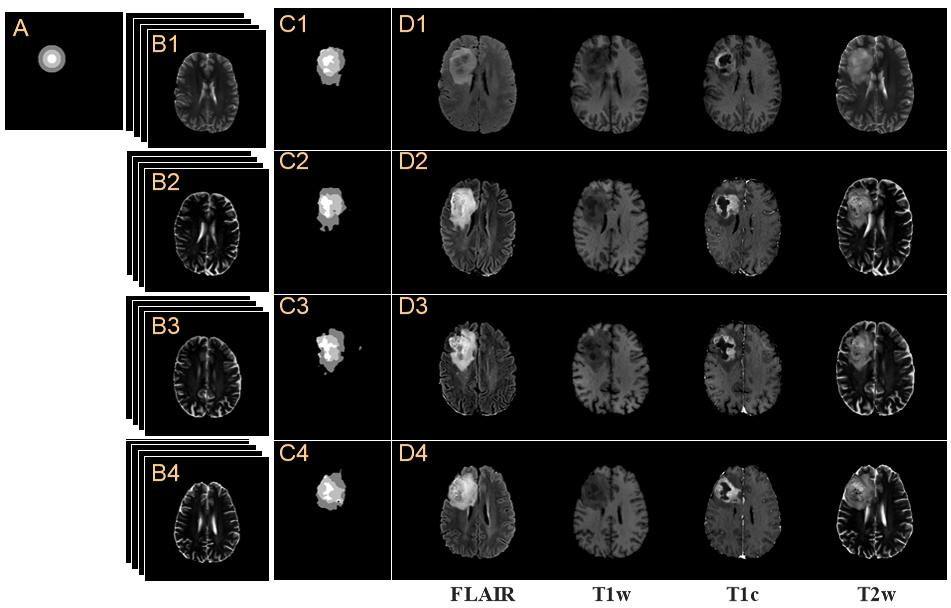}
  \caption{Synthesized brain tumor images, $\hat{\textbf{X}}_{T}$, from various normal brain images, $\textbf{X}_N$, with given concentric circles, $c$. (A) Given concentric circles and (B1-B4) various multi-contrast normal brain images, $\textbf{X}_N$. (C1-C4) $\hat{m}_{grade}$ generated from the given concentric circles for various normal brain images. (D) Synthesized multi-contrast brain tumor images, $\hat{\textbf{X}}_{T}$.}
  \label{fig:exp_samemask_variousbrain}
\end{figure*}

\begin{figure}[ht!]
\centering
  \includegraphics[width=0.7\linewidth]{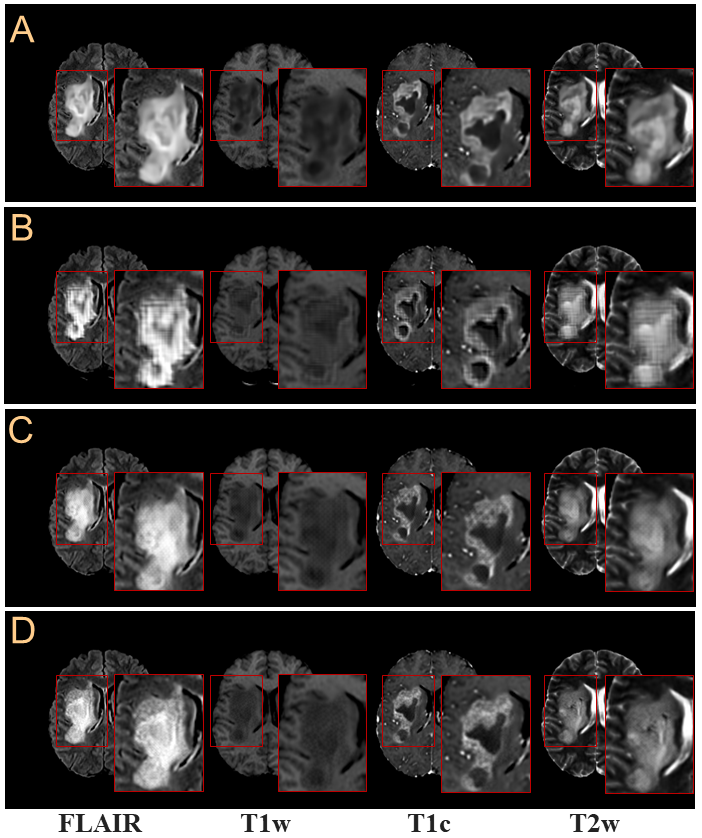}
  \caption{Synthesized brain tumor images from the proposed network, $G_{inpaint}$, trained by using various loss functions. The results according to (A) $\mathcal{L}_{inpaint:pixel}$, (B) $\mathcal{L}_{inpaint:pixel}+\mathcal{L}_{inpaint:adv}$, (C) $\mathcal{L}_{inpaint:pixel}+\mathcal{L}_{inpaint:cont}$, and (D)$\mathcal{L}_{inpaint:pixel}+\mathcal{L}_{inpaint:adv}+\mathcal{L}_{inpaint:cont}$. Tumor regions in red boxes were enlarged for clear comparison.}
  \label{fig:exp4loss_inpaint}
\end{figure}

We analyzed how the synthesized image changes depending on the size and position of the concentric circles with a given $\textbf{X}_N$ as shown in Fig.\ref{fig:exp_samebrain_variousmask}. The grade tumor masks are generated by $G_{grade}$ as shown in Fig.\ref{fig:exp_samebrain_variousmask}(C1-C4), generated with the given $\textbf{X}_N$ in Fig.\ref{fig:exp_samebrain_variousmask}(B) and various concentric circles in Fig.\ref{fig:exp_samebrain_variousmask}(A1-A4). The $\hat{m}_{grade}$ are generated according to the sizes, locations, and grade information of the concentric circles. In the cases of Fig.\ref{fig:exp_samebrain_variousmask}(A1) and (A2) having the same size and position but different grade information, the generated tumor masks in Fig.\ref{fig:exp_samebrain_variousmask}(C1) and (C2) have the same appearance, but only different grade information of $\hat{m}_{grade}$. In the cases of Fig.\ref{fig:exp_samebrain_variousmask}(A1) and (A3) having the same location but different sizes and grade information, $\hat{m}_{grade}$ in Fig.\ref{fig:exp_samebrain_variousmask}(C1) and (C3) are in the same location, but have different appearances and grade information. The concentric circles in Fig.\ref{fig:exp_samebrain_variousmask}(A1) and (A4) have the same size and same grade information, but different locations. The $\hat{m}_{grade}$ in Fig.\ref{fig:exp_samebrain_variousmask}(C1) and (C4) have the same size and grade information, but are different in appearance. These results show that the appearance of the tumor mask is affected by the size, grade information and location of the concentric circles. The images, $\hat{\textbf{X}}_T$, in Fig.\ref{fig:exp_samebrain_variousmask}(D1-D4) were synthesized through Fig.\ref{fig:exp_samebrain_variousmask}(A1-A4) and (B). The tumor parts of the synthesized brain tumor images depend on the concentric circles. It is proven that even one $\textbf{X}_N$ can be used to synthesize a variety of $\hat{\textbf{X}}_T$ if various concentric circles are used.

As shown in Fig.\ref{fig:exp_samemask_variousbrain}(A, B1-B4), only one example of concentric circles is combined with various normal brain images, $\textbf{X}_N$, as input of the proposed network. Various $\hat{m}_{grade}$ in Fig.\ref{fig:exp_samemask_variousbrain}(C1-C4) are generated depending on normal brain images while the same concentric circles are applied. In addition, the synthesized results, $\hat{\textbf{X}}_T$, show how the normal brain images affect the tumor-part images as shown in Fig.\ref{fig:exp_samemask_variousbrain}(D1-D4).

The pixel loss function, $\mathcal{L}_{inpaint:pix}$, is the simplest and most commonly used loss function in general deep learning algorithms. However, as shown in Fig.\ref{fig:exp4loss_inpaint}(A), the results tend to be blurry and less realistic. To overcome this problem, we include adversarial loss function, $\mathcal{L}_{inpaint:adv}$, and content loss function, $\mathcal{L}_{inpaint:cont}$, as shown in Fig.\ref{fig:exp4loss_inpaint}(B)-(D). When $\mathcal{L}_{inpaint:adv}$ is used together with $\mathcal{L}_{inpaint:pix}$ as shown in Fig.\ref{fig:exp4loss_inpaint}(B), a lot of lattice artifacts appear. In general, it is due to the instability of learning of the GAN-based algorithm. Here $\mathcal{L}_{inpaint:cont}$ is used together with $\mathcal{L}_{inpaint:pix}$ to make the synthesized images look more perceptually realistic as shown in Fig.\ref{fig:exp4loss_inpaint}(C). Finally, the results of the trained network using all the loss functions are shown in Fig.\ref{fig:exp4loss_inpaint}(D), where blurring and lattice artifacts are not seen.

\begin{figure*}[ht!]
\centering
  \includegraphics[width=0.8\linewidth]{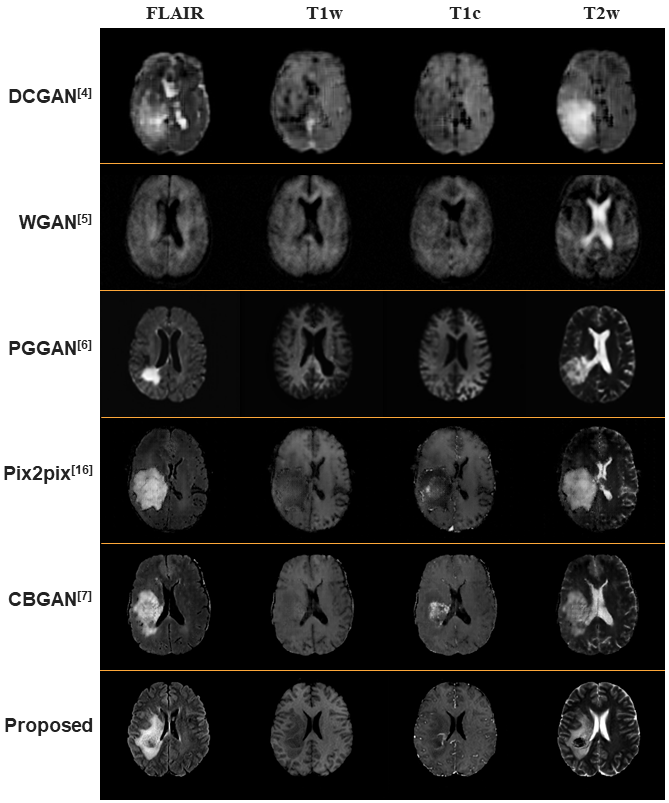}
  \caption{Comparison of brain tumor images synthesized by various augmentation methods.}
  \label{fig:exp_otherGAN}
\end{figure*}

\subsection{Comparisons with Other GAN Methods}

The results produced by the proposed method are compared with those produced by other GAN-based synthesis methods in Fig.\ref{fig:exp_otherGAN}. The first three rows are results of the unconditional GAN-based methods, which use random noise as an input, namely, DCGAN, WGAN, and PGGAN. As mentioned earlier, DCGAN is simple to train, but the results of DCGAN are unstable and unrealistic. The results generated by WGAN, on the other hand, are sharper, but unstable too. PGGAN is the best performance generative algorithm among the unconditional GANs, and brain tumor images are synthesized realistically. However, these unconditional GAN-based methods can only synthesize brain tumor images, not grade tumor masks, both of which are necessary for the training of supervised learning networks.

The 4th and 5th rows show the results of Pix2pix and CBGAN, respectively, where the brain tumor images were synthesized from the brain tumor label maps. Their results seem relatively unnatural because the normal parts must also be synthesized and the loss functions of their methods may not be sufficient. In the case of Pix2pix, there are lattice artifacts due to GAN's instability. In the case of CBGAN, similar to our proposed method, it uses various loss functions to remove lattice artifacts and to have learning stability. However, the non-tumor parts are displayed unnaturally compared to the proposed method. The proposed method fills the non-tumor part with the normal brain image as it is and generates only the tumor part. Thus, the results of the proposed method look most realistic.

\subsection{FID Score}
The realistic degree of brain tumor images is quantitatively analyzed through the Fréchet inception distance (FID) score \cite{heusel2017gans}, which has generally been used to measure the performance of GAN-based algorithms. The FID score represents the distance between training data and generated data. A lower FID score indicates greater similarity with training data. Therefore, as listed in Table \ref{table:FID_score}, the FID score is used to measure the similarity between the training data and the generated data according to various loss functions, and the FID scores are compared for the proposed method and other GAN-based algorithms, CBGAN and Pix2pix. The proposed method had lower FID values than the CBGAN and Pix2pix methods because our method uses real normal brain images for the non-tumor part.

\begin{table*}[htbp]
\centering
\caption{FID (Fréchet Inception Distance) score}
\label{table:FID_score} 
\begin{adjustbox}{max width=\linewidth} 
\begin{tabular}{@{}lccccc@{}}
\toprule[1pt]\midrule[0.3pt]
\multicolumn{1}{c}{\multirow{2}{*}{Method}} & \multicolumn{5}{c}{Contrast}                                                                                                              \\ \cmidrule(){2-6} 
\multicolumn{1}{c}{}                        & \multicolumn{1}{c}{FLAIR} & \multicolumn{1}{c}{T1w} & \multicolumn{1}{c}{T1c} & \multicolumn{1}{c}{T2} & \multicolumn{1}{c}{\textbf{Avg}} \\ \toprule
\textit{$\mathcal{L}_{Inpaint}^{pixel}$}    & 0.8896                    & 0.2121                  & 0.6075                  & 4.8519                 & 0.1.6403                         \\[2pt]
\textit{$\mathcal{L}_{Inpaint}^{pixel},\mathcal{L}_{Inpaint}^{adv}$} & 0.7579   & 0.1707          & \textbf{0.4905}         & 0.3.9505               & 0.1.3424                         \\[2pt]
\textit{\textbf{$\mathcal{L}_{Inpaint}^{pixel},\mathcal{L}_{Inpaint}^{adv},\mathcal{L}_{Inpaint}^{cont}$}} & \textbf{0.5273}& \textbf{1.2095}        & 0.4306   &\textbf{2.4712}         & \textbf{1.1597}  \\ [2pt]
\textit{CBGAN}\cite{mok2018learning}                                       & 1.4406               & 0.4358                  & 0.5590                 & 4.4123                 & 1.712                      \\[2pt]
\textit{Pix2pix}\cite{isola2017image}                                       & 3.1472              & 1.3201                  & 2.2865                 & 13.4541                & 5.052                      \\[2pt]
\midrule[0.3pt]\bottomrule[1pt]
\end{tabular}
\end{adjustbox}
\end{table*}

\subsection{Segmentation Performance through Data Augmentation}

In tumor segmentation experiment, three metrics of dice, sensitivity, and precision are usually used to measure the performance. Dice, sensitivity, and precision are each defined as follows:

\begin{equation}
\begin{aligned}
\centering
&Dice=\frac{2 \cdot TP}{2 \cdot TP + FP + FN},\\
&Sensitivity=\frac{TP}{TP+FN},\\ 
&Precision=\frac{TP}{TP+FP}
\end{aligned}
\end{equation}
where TP, FP, FN, and TN denotes true positive, false positive, false negative and true negative, respectively.

Dice is affected by FN as well as FP. Therefore, it can be used to comprehensively analyze the performance. However, dice cannot provide enough information to analyze whether the low dice is due to FP or FN. Therefore, analyses of sensitivity and precision are necessary. Sensitivity is affected by FN, and precision is affected by FP.
 
UNet \cite{ronneberger2015u} was used for tumor segmentation. Only the numbers of input channels and output channels of UNet were modified in the experiments. The input channel has multi-contrast MR images of the brain tumor, and the output channel has five segmented images of the whole tumor (WT), tumor core (TC), enhancing tumor (ET), non-tumor part, and background. We measured the segmentation performance at epochs of 200 and used the Adam optimizer with a learning rate of 0.0002.
 
The experimental results for brain tumor segmentation are listed in Table \ref{table:segmentation-performance}. A total of 5k brain tumor images were randomly chosen from a total of 12.5k. The UNet for brain tumor segmentation was trained by using the selected 5k brain tumor images and the augmented 7.5k brain tumor images that were synthesized by various augmentation methods. In each case, the segmentation performance of WT, TC, and ET of the tumors was measured using the metrics of dice, sensitivity, and precision. The proposed method was compared with conventional augmentation methods and other GAN-based methods. Furthermore, for the proposed method, comparative experiments were performed according to the loss functions.

The results of the conventional augmentation methods show a meaningful performance improvement over the basic performance from only the real 5k brain tumor images. This is why many algorithms have adopted data augmentation for network training. We analyzed the performance depending on each loss function and proved the validity of the loss function. The proposed method shows a significant performance improvement over the conventional augmentation methods and other GAN-based methods.

\begin{table*}[h!]
\centering
\caption{Segmentation Performances}
\label{table:segmentation-performance} 
\begin{adjustbox}{max width=\textwidth} 
\begin{tabular}{@{}l l cc ccc ccc ccc@{}}
\toprule[1pt]\midrule[0.3pt]
\multirow{2}{*}{} & \multirow{2}{*}{Augment Method} & \multicolumn{2}{c}{ \# Training Images}       & \multicolumn{3}{c}{Dice}      & \multicolumn{3}{c}{Sensitivity}       & \multicolumn{3}{c}{Precision}     \\
\cmidrule(lr){3-4} \cmidrule(lr){5-7} \cmidrule(lr){8-10} \cmidrule(lr){11-13} 
                  &                                         & Real      & Synthesized               & WT            & TC            & ET            & WT            & TC            & ET            & WT            & TC            & ET          \\  \toprule
Real Images              & $\cdot$                          & 5k        & $\cdot$                   & 0.723         & 0.612         & 0.346         & 0.673         & 0.600         & 0.323         & 0.816         & 0.657         & 0.442       \\
                  & $\cdot$                                 & 12.5k     & $\cdot$                   & 0.743         & 0.648         & 0.463         & 0.670         & 0.639         & 0.357         & 0.868         & 0.737         & 0.551       \\  \midrule 
Conventional Methods  & \textit{Rotate}                     & 5k        & 7.5k                      & 0.734         & 0.610         & 0.338         & 0.694         & 0.569         & 0.298         & 0.813         & 0.699         & 0.466       \\
                  & \textit{Flip}                           & 5k        & 7.5k                      & 0.731         & 0.549         & 0.305         &\textbf{0.700} & 0.551         & 0.278         & 0.797         & 0.581         & 0.416       \\
                  & \textit{Brightness, color jittering}    & 5k        & 7.5k                      &\textbf{0.738} & 0.621         & 0.369         & 0.679         & 0.579         & 0.334         & 0.833         & 0.701         & 0.480       \\
                  & \textit{All of conventional methods} & 5k & 7.5k                  & 0.735         & 0.619         & 0.351         & 0.691         & \textbf{0.621}         & 0.322         & 0.816         & 0.650         & 0.453       \\
\textbf{Proposed Methods}& \textit{$\mathcal{L}_{Inpaint}^{pixel}$}& 5k        & 7.5k               & 0.736         & 0.622         & 0.366         & 0.682         & 0.611         & 0.330         & 0.829         & 0.670         & 0.476       \\
    & \textit{$\mathcal{L}_{Inpaint}^{pixel},\mathcal{L}_{Inpaint}^{adv}$} & 5k  & 7.5k             & 0.730         & 0.628         & 0.374         & 0.671         & 0.600         & 0.340         & 0.831         & 0.699         & 0.478       \\
    & \textbf{$\mathcal{L}_{Inpaint}^{pixel},\mathcal{L}_{Inpaint}^{adv},\mathcal{L}_{Inpaint}^{cont}$} & 5k  & 7.5k    &0.737 &\textbf{0.646} &\textbf{0.390} & 0.663     &0.619   & 0.334         &\textbf{0.864} &\textbf{0.706} &\textbf{0.536} \\
GAN-based Methods        & \textit{CBGAN} \cite{mok2018learning}    & 5k        & 7.5k             & 0.736         & 0.588         & 0.386         & 0.679         & 0.550         &\textbf{0.343} & 0.854         & 0.670         & 0.507     \\ 
                        & \textit{Pix2pix} \cite{isola2017image}    & 5k        & 7.5k             & 0.737         & 0.640         & 0.354         & 0.686         & 0.619         &0.314 & 0.831         & 0.688         & 0.476     \\ 
\midrule[0.3pt]\bottomrule[1pt]
\end{tabular}
\end{adjustbox}
\end{table*}

\section{Discussion}
\label{sec:discussion}

Three metric values of the proposed method are high compared to those of conventional augmentation methods and other GAN-based methods. Although the conventional augmentation methods improved the segmentation performance, the improvement was not sufficient. The other GAN-based methods of CBGAN and pix2pix also improved the segmentation performance. They synthesized not only the tumor area but also non-tumor parts. Therefore, the synthesized brain tumor images may be unrealistic and thus cannot improve the segmentation performance significantly. Because our proposed method fills the non-tumor part with the synthesized tumor and uses normal brain images for non-tumor parts in the synthesized brain tumor images, it produces more realistic images. In other words, when the data augmented by our method are used for training of the segmentation network, the dice value of the proposed method, which is the combined value of precision and sensitivity, is improved.

There are several considerations regarding our proposed method. The first consideration is the absence of pathological knowledge of the actual tumor location. Although we have freely determined the location, size and grade information of the tumor, it is necessary to examine whether the synthesized tumors are pathologically realistic shape. Combining our algorithm with pathological knowledge can improve the data augmentation performance. The next consideration is the deformation of the surrounding normal tissue due to the tumor. Because the non-tumor part is taken from a normal brain image as it is, the deformation of the surrounding area is not shown. If all of these are taken into account, it is expected that it would be possible to synthesize not only photographically realistic but also pathologically realistic brain tumor images.

\section{Conclusion}
\label{sec:conclusion}
Our proposed method simplifies grade tumor masks into the concentric circles, so that the sizes and location of the concentric circles can be easily controlled to provide diversity of tumor images. New grade tumor masks can be stably generated from concentric circles, and brain tumor images are synthesized through deep neural networks. The five variables defining three concentric circles are very simple to manipulate, and the number of healthy people is much larger than that of brain tumor patients. This can enable the synthesis of numerous brain tumor images. In terms of data augmentation, the proposed method can successfully synthesize brain tumor images that can be used to train tumor segmentation networks or other deep neural networks.


{
\footnotesize
\bibliographystyle{unsrt}
\bibliography{ref}

\begin{thebibliography}{10}

\bibitem{menze2014multimodal}
Bjoern~H Menze, Andras Jakab, Stefan Bauer, Jayashree Kalpathy-Cramer, Keyvan
  Farahani, Justin Kirby, Yuliya Burren, Nicole Porz, Johannes Slotboom, Roland
  Wiest, et~al.
\newblock The multimodal brain tumor image segmentation benchmark (brats).
\newblock {\em IEEE transactions on medical imaging}, 34(10):1993--2024, 2014.

\bibitem{van2013wu}
David~C Van~Essen, Stephen~M Smith, Deanna~M Barch, Timothy~EJ Behrens, Essa
  Yacoub, Kamil Ugurbil, Wu-Minn~HCP Consortium, et~al.
\newblock The wu-minn human connectome project: an overview.
\newblock {\em Neuroimage}, 80:62--79, 2013.

\bibitem{eaton2018improving}
Zach Eaton-Rosen, Felix Bragman, Sebastien Ourselin, and M~Jorge Cardoso.
\newblock Improving data augmentation for medical image segmentation.
\newblock 2018.

\bibitem{radford2015unsupervised}
Alec Radford, Luke Metz, and Soumith Chintala.
\newblock Unsupervised representation learning with deep convolutional
  generative adversarial networks.
\newblock {\em arXiv preprint arXiv:1511.06434}, 2015.

\bibitem{arjovsky2017wasserstein}
Martin Arjovsky, Soumith Chintala, and L{\'e}on Bottou.
\newblock Wasserstein gan.
\newblock {\em arXiv preprint arXiv:1701.07875}, 2017.

\bibitem{karras2017progressive}
Tero Karras, Timo Aila, Samuli Laine, and Jaakko Lehtinen.
\newblock Progressive growing of gans for improved quality, stability, and
  variation.
\newblock {\em arXiv preprint arXiv:1710.10196}, 2017.

\bibitem{mok2018learning}
Tony~CW Mok and Albert~CS Chung.
\newblock Learning data augmentation for brain tumor segmentation with
  coarse-to-fine generative adversarial networks.
\newblock In {\em International MICCAI Brainlesion Workshop}, pages 70--80.
  Springer, 2018.

\bibitem{shin2018medical}
Hoo-Chang Shin, Neil~A Tenenholtz, Jameson~K Rogers, Christopher~G Schwarz,
  Matthew~L Senjem, Jeffrey~L Gunter, Katherine~P Andriole, and Mark Michalski.
\newblock Medical image synthesis for data augmentation and anonymization using
  generative adversarial networks.
\newblock In {\em International Workshop on Simulation and Synthesis in Medical
  Imaging}, pages 1--11. Springer, 2018.

\bibitem{goodfellow2014generative}
Ian Goodfellow, Jean Pouget-Abadie, Mehdi Mirza, Bing Xu, David Warde-Farley,
  Sherjil Ozair, Aaron Courville, and Yoshua Bengio.
\newblock Generative adversarial nets.
\newblock In {\em Advances in neural information processing systems}, pages
  2672--2680, 2014.

\bibitem{ronneberger2015u}
Olaf Ronneberger, Philipp Fischer, and Thomas Brox.
\newblock U-net: Convolutional networks for biomedical image segmentation.
\newblock In {\em International Conference on Medical image computing and
  computer-assisted intervention}, pages 234--241. Springer, 2015.

\bibitem{pereira2016brain}
S{\'e}rgio Pereira, Adriano Pinto, Victor Alves, and Carlos~A Silva.
\newblock Brain tumor segmentation using convolutional neural networks in mri
  images.
\newblock {\em IEEE transactions on medical imaging}, 35(5):1240--1251, 2016.

\bibitem{havaei2017brain}
Mohammad Havaei, Axel Davy, David Warde-Farley, Antoine Biard, Aaron Courville,
  Yoshua Bengio, Chris Pal, Pierre-Marc Jodoin, and Hugo Larochelle.
\newblock Brain tumor segmentation with deep neural networks.
\newblock {\em Medical image analysis}, 35:18--31, 2017.

\bibitem{bermudez2018learning}
Camilo Bermudez, Andrew~J Plassard, Larry~T Davis, Allen~T Newton, Susan~M
  Resnick, and Bennett~A Landman.
\newblock Learning implicit brain mri manifolds with deep learning.
\newblock In {\em Medical Imaging 2018: Image Processing}, volume 10574, page
  105741L. International Society for Optics and Photonics, 2018.

\bibitem{han2018gan}
Changhee Han, Hideaki Hayashi, Leonardo Rundo, Ryosuke Araki, Wataru Shimoda,
  Shinichi Muramatsu, Yujiro Furukawa, Giancarlo Mauri, and Hideki Nakayama.
\newblock Gan-based synthetic brain mr image generation.
\newblock In {\em 2018 IEEE 15th International Symposium on Biomedical Imaging
  (ISBI 2018)}, pages 734--738. IEEE, 2018.

\bibitem{bowles2018gan}
Christopher Bowles, Liang Chen, Ricardo Guerrero, Paul Bentley, Roger Gunn,
  Alexander Hammers, David~Alexander Dickie, Maria~Vald{\'e}s Hern{\'a}ndez,
  Joanna Wardlaw, and Daniel Rueckert.
\newblock Gan augmentation: augmenting training data using generative
  adversarial networks.
\newblock {\em arXiv preprint arXiv:1810.10863}, 2018.

\bibitem{isola2017image}
Phillip Isola, Jun-Yan Zhu, Tinghui Zhou, and Alexei~A Efros.
\newblock Image-to-image translation with conditional adversarial networks.
\newblock In {\em Proceedings of the IEEE conference on computer vision and
  pattern recognition}, pages 1125--1134, 2017.

\bibitem{yu2018generative}
Jiahui Yu, Zhe Lin, Jimei Yang, Xiaohui Shen, Xin Lu, and Thomas~S Huang.
\newblock Generative image inpainting with contextual attention.
\newblock In {\em Proceedings of the IEEE Conference on Computer Vision and
  Pattern Recognition}, pages 5505--5514, 2018.

\bibitem{liu2018image}
Guilin Liu, Fitsum~A Reda, Kevin~J Shih, Ting-Chun Wang, Andrew Tao, and Bryan
  Catanzaro.
\newblock Image inpainting for irregular holes using partial convolutions.
\newblock In {\em Proceedings of the European Conference on Computer Vision
  (ECCV)}, pages 85--100, 2018.

\bibitem{iizuka2017globally}
Satoshi Iizuka, Edgar Simo-Serra, and Hiroshi Ishikawa.
\newblock Globally and locally consistent image completion.
\newblock {\em ACM Transactions on Graphics (ToG)}, 36(4):107, 2017.

\bibitem{yeh2017semantic}
Raymond~A Yeh, Chen Chen, Teck Yian~Lim, Alexander~G Schwing, Mark
  Hasegawa-Johnson, and Minh~N Do.
\newblock Semantic image inpainting with deep generative models.
\newblock In {\em Proceedings of the IEEE Conference on Computer Vision and
  Pattern Recognition}, pages 5485--5493, 2017.

\bibitem{zhao2019guided}
Yinan Zhao, Brian Price, Scott Cohen, and Danna Gurari.
\newblock Guided image inpainting: Replacing an image region by pulling content
  from another image.
\newblock In {\em 2019 IEEE Winter Conference on Applications of Computer
  Vision (WACV)}, pages 1514--1523. IEEE, 2019.

\bibitem{yu2019free}
Jiahui Yu, Zhe Lin, Jimei Yang, Xiaohui Shen, Xin Lu, and Thomas~S Huang.
\newblock Free-form image inpainting with gated convolution.
\newblock In {\em Proceedings of the IEEE International Conference on Computer
  Vision}, pages 4471--4480, 2019.

\bibitem{park2019semantic}
Taesung Park, Ming-Yu Liu, Ting-Chun Wang, and Jun-Yan Zhu.
\newblock Semantic image synthesis with spatially-adaptive normalization.
\newblock In {\em Proceedings of the IEEE Conference on Computer Vision and
  Pattern Recognition}, pages 2337--2346, 2019.

\bibitem{simonyan2014very}
Karen Simonyan and Andrew Zisserman.
\newblock Very deep convolutional networks for large-scale image recognition.
\newblock {\em arXiv preprint arXiv:1409.1556}, 2014.

\bibitem{gatys2016image}
Leon~A Gatys, Alexander~S Ecker, and Matthias Bethge.
\newblock Image style transfer using convolutional neural networks.
\newblock In {\em Proceedings of the IEEE conference on computer vision and
  pattern recognition}, pages 2414--2423, 2016.

\bibitem{johnson2016perceptual}
Justin Johnson, Alexandre Alahi, and Li~Fei-Fei.
\newblock Perceptual losses for real-time style transfer and super-resolution.
\newblock In {\em European conference on computer vision}, pages 694--711.
  Springer, 2016.

\bibitem{ulyanov2016instance}
Dmitry Ulyanov, Andrea Vedaldi, and Victor Lempitsky.
\newblock Instance normalization: The missing ingredient for fast stylization.
\newblock {\em arXiv preprint arXiv:1607.08022}, 2016.

\bibitem{kingma2014adam}
Diederik~P Kingma and Jimmy Ba.
\newblock Adam: A method for stochastic optimization.
\newblock {\em arXiv preprint arXiv:1412.6980}, 2014.

\bibitem{heusel2017gans}
Martin Heusel, Hubert Ramsauer, Thomas Unterthiner, Bernhard Nessler, and Sepp
  Hochreiter.
\newblock Gans trained by a two time-scale update rule converge to a local nash
  equilibrium.
\newblock In {\em Advances in Neural Information Processing Systems}, pages
  6626--6637, 2017.

\end{thebibliography}
}

\end{document}